\documentclass[a4paper,12pt, epsfig]{article}
\usepackage{epsfig}
\usepackage{amssymb}
\usepackage{amsfonts}

\newskip\humongous \humongous=0pt plus 1000pt minus 1000pt

\newif\ifdtup

\catcode`\@=11

\@addtoreset{equation}{section}
\def\theequation{\thesection.\arabic{equation}}

\def\@normalsize{\@setsize\normalsize{15pt}\xiipt\@xiipt
\abovedisplayskip 14pt plus3pt minus3pt%
\belowdisplayskip \abovedisplayskip
\abovedisplayshortskip \z@ plus3pt%
\belowdisplayshortskip 7pt plus3.5pt minus0pt}

\def\small{\@setsize\small{13.6pt}\xipt\@xipt
\abovedisplayskip 13pt plus3pt minus3pt%
\belowdisplayskip \abovedisplayskip
\abovedisplayshortskip \z@ plus3pt%
\belowdisplayshortskip 7pt plus3.5pt minus0pt
\def\@listi{\parsep 4.5pt plus 2pt minus 1pt
      \itemsep \parsep
      \topsep 9pt plus 3pt minus 3pt}}

\relax

\catcode`@=12

\catcode`\@=11

\def\section{\@startsection{section}{1}{\z@}{3.5ex plus 1ex minus
    .2ex}{2.3ex plus .2ex}{\large\bf}}

\def\thesection{\arabic{section}}
\def\thesubsection{\arabic{section}.\arabic{subsection}}

\def\appendix{\setcounter{section}{0}
  \def\thesection{Appendix \Alph{section}}
  \def\thesubsection{\Alph{section}.\arabic{subsection}}
  \def\theequation{\Alph{section}.\arabic{equation}}}

\def\SymBoxes#1#2#3#4{\newdimen\un@t \un@t#3%
\raisebox{#1}{\rule{#2\un@t}{#4}\hskip-#2\un@t
\@tempdimb\un@t \advance\@tempdimb by-#4\@tempcntb#2\relax%
\@whilenum{\@tempcntb>0}\do{
\rule{#4}{\un@t}\hskip\@tempdimb \advance\@tempcntb by\m@ne}%
\hskip-#2\un@t \rule[\un@t]{#2\un@t}{#4}%
\rule[\un@t]{#4}{#4}\hskip-#4
\rule{#4}{\un@t}}\hskip-#4}                

\begin{document}

\newcommand{\beq}{\begin{equation}}
\newcommand{\eeq}{\end{equation}}
\newcommand{\bea}{\begin{eqnarray}}
\newcommand{\eea}{\end{eqnarray}}
\newcommand{\beas}{\begin{eqnarray*}}
\newcommand{\eeas}{\end{eqnarray*}}
\newcommand{\defi}{\stackrel{\rm def}{=}}
\newcommand{\non}{\nonumber}
\newcommand{\bquo}{\begin{quote}}
\newcommand{\enqu}{\end{quote}}
\renewcommand{\(}{\begin{equation}}
\renewcommand{\)}{\end{equation}}
\def\de{\partial}
\def\Om{\ensuremath{\Omega}}
\def\Tr{ \hbox{\rm Tr}}
\def\H{ \hbox{\rm H}}
\def\HE{ \hbox{$\rm H^{even}$}}
\def\HO{ \hbox{$\rm H^{odd}$}}
\def\HEO{ \hbox{$\rm H^{even/odd}$}}
\def\HOE{ \hbox{$\rm H^{odd/even}$}}
\def\HHO{ \hbox{$\rm H_H^{odd}$}}
\def\HHEO{ \hbox{$\rm H_H^{even/odd}$}}
\def\HHOE{ \hbox{$\rm H_H^{odd/even}$}}
\def\K{ \hbox{\rm K}}
\def\Im{ \hbox{\rm Im}}
\def\Ker{ \hbox{\rm Ker}}
\def\const{\hbox {\rm const.}}
\def\o{\over}
\def\im{\hbox{\rm Im}}
\def\re{\hbox{\rm Re}}
\def\bra{\langle}\def\ket{\rangle}
\def\Arg{\hbox {\rm Arg}}
\def\Re{\hbox {\rm Re}}
\def\Im{\hbox {\rm Im}}
\def\exo{\hbox {\rm exp}}
\def\diag{\hbox{\rm diag}}
\def\longvert{{\rule[-2mm]{0.1mm}{7mm}}\,}
\def\a{\alpha}
\def\dag{{}^{\dagger}}
\def\tq{{\widetilde q}}
\def\p{{}^{\prime}}
\def\W{W}
\def\N{{\cal N}}
\def\hsp{,\hspace{.7cm}}
\def\bo{\ensuremath{\hat{b}_1}}
\def\bfo{\ensuremath{\hat{b}_4}}
\def\co{\ensuremath{\hat{c}_1}}
\def\cfo{\ensuremath{\hat{c}_4}}
\newcommand{\C}{\ensuremath{\mathbb C}}
\newcommand{\Z}{\ensuremath{\mathbb Z}}
\newcommand{\R}{\ensuremath{\mathbb R}}
\newcommand{\rp}{\ensuremath{\mathbb {RP}}}
\newcommand{\cp}{\ensuremath{\mathbb {CP}}}
\newcommand{\vac}{\ensuremath{|0\rangle}}
\newcommand{\vact}{\ensuremath{|00\rangle}                    }
\newcommand{\oc}{\ensuremath{\overline{c}}}
\begin{titlepage}
\begin{flushright}
SISSA 39/2008/EP
\end{flushright}
\bigskip
\def\thefootnote{\fnsymbol{footnote}}

\begin{center}
{\large {\bf
Geometric Engineering 5d Black Holes with Rod Diagrams
  } }
\end{center}

\bigskip
\begin{center}
{\large  Jarah Evslin\footnote{\texttt{evslin@sissa.it}}}
\end{center}

\renewcommand{\thefootnote}{\arabic{footnote}}

\begin{center}
\vspace{1em}
{\em  { SISSA,\\
Via Beirut 2-4,\\
I-34014, Trieste, Italy}

}
\end{center}

\noindent
\begin{center} {\bf Abstract} \end{center}
Static solutions of 5-dimensional gravity with two spatial Killing vectors are characterized by their rod structures.  In this note we describe how the orbifold singularities and the topologies of the horizons and asymptotic regions can be determined from the corresponding rod diagrams.  As an example we introduce the black lens, a static 5-dimensional black hole with a horizon of lens space topology which is asymptotically Minkowski space.  The solution is novel in that the asymptotic Minkowski space is not quotiented.  However it suffers from a naked singularity.  While the conical and orbifold singularities have been removed, two spherical curvature singularities remain.  These singularities do not contribute to the ADM mass, and the thermodynamics of the black lens is well behaved, although its entropy is lower than that of a Tangherlini black hole of the same mass.


\vfill

\begin{flushleft}
{\today}
\end{flushleft}
\end{titlepage}

\hfill{}


\setcounter{footnote}{0}
\section{Introduction}
In higher-dimensional general relativity it is easy to make a black hole whose event horizon has a topology that is not a product of spheres.  For example, start with Schwarzschild-Tangherlini in 5-dimensions, whose horizon has topology $S^3$, and quotient by a free discrete symmetry of the sphere, like a cyclic group or a nonabelian crystallographic group.  Then the horizon has the quotient topology, as does the asymptotic region.  

But what if one doesn't want to quotient the asymptotic sphere, and instead one imposes that the configuration be asymptotically globally Minkowski?  In this case, until this century, in all known solutions in all numbers of dimensions the horizon had spherical topology.  This changed with the discovery of the black ring solution in Ref.~\cite{ring}, it was found that horizon topologies may be products of spheres.  This again changed with the discovery of the black Saturn \cite{saturn} and then the di-ring \cite{di,di2,di3} and bi-ring \cite{bi,bi2}, where it was found that horizon topologies may be unions of products of spheres.  Already at this stage, the inverse scattering transformation used to obtain the solutions requires the inversion of a matrix, called $\Gamma$.   While $\Gamma$ is always nondegenerate for the singly-rotating black Saturn, it is not known whether it degenerates for the di-ring and bi-ring and so it is not known whether these solutions posses naked singularities.  Instead authors have contented themselves with the fact that thermodynamic quantities are positive and so there are no apparent pathologies at least at the horizon.  It has been argued that all of these configurations except for the black hole, black ring and black Saturn are also thermodynamically unstable as different black objects will have different temperatures and so cannot be in a stable equilibrium.  While black Saturn is also thermodynamically unstable, it is metastable in a small window \cite{metastable}.  In extensions of general relativity many more possibilities appear \cite{simp1,simp2,simp3,Yaz1,Yaz2,Yaz3} and it would be interesting to see if our results extend to these cases.

In the present note, we provide a modest extension of the known topologies of event horizons of 5d black holes, resulting in countably infinite hair. We provide an infinite class of new static solutions, called black lenses, with horizon topology $L(p^2+1,1)$, which is the quotient of the 3-sphere by the free cyclic group of order $p^2+1$ for any integer $p>1$.  These are not composite, and so are not thermodynamically unstable in the above sense.  However they may well be classically unstable, and we will indeed find that their entropies are less than that of a Schwarzschild-Tangherlini black hole with the same mass.  The transformation matrix $\Gamma$ indeed degenerates in our solutions, on two spheres that lie between the horizon and infinity.  These spheres appear to be timelike curvature singularities, as the Riemann tensor squared suffers a $1/r^6$ divergence.  Under a deformation of the solution the spheres disappear, and so we hope that in the rotating or quantum case they will not be present.  As in the di-ring and bi-ring case, the thermodynamic quantities that we calculate show no signs of a pathology and the Komar mass of the horizon is equal to the ADM mass, so there is no contribution from the singularities.  In deriving the solution there are also conical singularities which end on orbifold singularities, but we are able to eliminate both by correctly choosing the parameters in our ans\"atz.

At intermediate distances beyond the horizon, the spacetime appears not to be Minkowski, but is more similar to an ALE of multicentered Taub-NUT space, in which the existence of black holes with lens space topology is well-known (see Ref.~\cite{lens} and references therein for some examples).  However using Harmark's rod description of general relativity with commuting isometries \cite{Harmark} we are able to engineer a spacetime that is asymptotically globally Minkowski.  This suggests that the same strategy may be applied to other configurations which are known to exist only in asymptotic Taub-NUT or quotient spaces, such as multiple black holes of spherical topology.  One may even conjecture that the set of allowed horizon topologies is independent of the asymptotics.  This would imply that the quotient of the sphere by a crystallographic group is consistent with asymptotic Minkowski space.  As the crystallographic group does not commute with the $U(1)^2$ isometry group of a spatial slice, this would provide a counterexample to the conjecture that stationary solutions of 5d general relativity have at least two spacelike isometries.

In Sec.~\ref{topseca} we describe how one can read the topology of the horizon and the asymptotic region, as well as the orbifold singularities, from the rod structure of a 5-dimensional stationary solution to Einstein's equations with vanishing stress tensor.  Conversely, we see how one may choose a rod structure to engineer the desired topology.  Next in Sec.~\ref{asec} we consider a particular example, the singular black lens.  We choose a certain rod structure which, according to the arguments of the previous section, corresponds to a single black hole with lens space horizon and asymptotes to globally Minkowski space and we find the metric as a function of several parameters.  These parameters are fixed in Sec.~\ref{parasec} when we demand that the coordinates asymptote to the usual Cartesian coordinates and that the spacetime be free of conical singularities.  Fixing the topology of the lens space horizon, we are left with a single free parameter.  Finally in Sec.~\ref{propsec} we see that the remaining parameter corresponds to an overall scale.  We fix it by fixing the ADM mass, and we calculate the entropy and temperature.  These quantities are found to be positive and to satisfy the 5d Smarr relation.

\section{Topology from rod structures} \label{topseca}

\subsection{Rod structures}

Einstein's equations are difficult to solve in 5-dimensions.  Therefore we will restrict our attention to stationary spacetimes with two spacelike Killing vectors.  As the metric is independent of the Killing directions, it depends on only two variables and Einstein's equations reduce to a well-studied 2-dimensional integrable system.  A classification of such solutions was provided by Harmark in Ref.~\cite{Harmark}.  He found that the configuration is characterized by the degenerations of the orbits of the Killing vectors.  Moreover, all of the degenerations occur on a single infinite line with coordinate $z$, and, up to a scale, only one Killing vector out of the 3-dimensional space of Killing vectors vanishes at each point.  To make things even simpler, the vector that vanishes is constant on open intervals along the line, it only changes at a finite set $\{ a_i \}$ of points on the line.  These points partition the line into intervals called rods.  

Thus a configuration is entirely determined by a set of points $\{a_i\}$ and Killing vectors $\{v_i\}$.  In general there will be a conical singularity at a rod if the period of the vanishing Killing coordinate is not $2\pi$ times the distance $\rho$ from the rod.  If there are two rods in a given spatial direction, then as we will see in Sec.~\ref{parasec} the elimination of the conical singularity on one rod fixes the period and the elimination on the other fixes one of the parameters.  More generally when rod vectors are linearly dependent, each relation on the rod vectors leads to a constraint on the parameters.  Rods on which $v_i$ is timelike correspond to horizons, and the period of the imaginary part of this Killing direction gives the inverse temperature.  When there are multiple horizons, the imaginary part of the periods may be set to be equal, as in the case of spatial rods.  In this case the equality is not necessary for the cancellation of conical singularities, but rather for the thermal equilibrium of the various components.  However in the present note we will only have one timelike rod.

In general there are also orbifold singularities at the intersections of spacelike rods at $a_i$.  These occur when the symplectic intersection product $q$ of the cycles that degenerate on the two sides is not equal to $\pm 1$, because in this case there will be a $\Z_q$ valued cycle that does not degenerate at $a_i$.  We will see shortly that in this case the link of $a_i$ will not be a $S^3$ but rather $L(q,1)$.  This identifies $a_i$ as a $\Z_q$ orbifold singularity.  Therefore we will need to choose the vectors of our rods such that the intersection product of the collapsing Killing circles is always equal to $\pm 1$.

The simplest example of a rod structure in 5-dimensions is that of flat space.  It has two rods, one at $z<0$ with vector $v_0=(t,\phi,\psi)=(0,1,0)$ and one at $z>0$ with vector $v_1=(0,0,1)$.  If one thinks of each spatial slice as $(z_1,z_2)\in\C^2$, then $\phi$ and $\psi$ are just the phases of $z_1$ and $z_2$ respectively.  Then one understands that on the left rod $\phi$ degenerates because $z_1=0$ and on the right rod $\psi$ degenerates because $z_2=0$.  At the origin there is no orbifold singularity because the intersection product of the vanishing vectors is equal to $\pm 1$
\beq
(v_0,v_1)=v_0^\phi v_1^\psi-v_0^\psi v_1^\phi=1. \label{symp}
\eeq

The condition (\ref{symp}) is invariant under symplectic transformations of the $v_i$, which are just SL($2,\Z$) large diffeomorphisms of the spacelike isometry torus.  For example one may instead set $v_1=(0,p,1)$ for any value of $p$ and the product will be unchanged.  Of course, one needs to ensure not only that the orbifold singularity is gone but also that there is no conical singularity, which is most likely impossible in the present case if $\phi$ and $\psi$ are chosen to be orthogonal, as the period of the $v_1$ circle will invariably be greater than that of the $v_0$ circle.  In our solution we will find such an intersection and, while $\phi$ and $\psi$ are orthogonal in the asymptotic regime, they will not be orthogonal at the intersection and so the conical singularities can be consistently eliminated on both sides.

A black hole is a solution with an event horizon, which is roughly a surface where the time component of the metric vanishes.  In the rod language, it corresponds to the degeneration of a timelike Killing vector.  Combining a black hole with flat space one arrives at a black hole in asymptotically flat space, which is described by three rods with vectors
\beq
v_0=(0,1,0)\hsp v_1=(1,0,0)\hsp v_2=(0,0,1). \label{sfera}
\eeq   
The horizon lies on the compact rod $v_1$, but the asymptotic region only depends on the semi-infinite rods $v_0$ and $v_2$ and so the configuration is asymptotically Minkowski.  To calculate the topology of the horizon, note that the horizon corresponds to the finite rod, which is a finite interval with two spacelike Killing circles.  Thus it is a fibration of $T^2$ over an interval.  It degenerates at the two ends of the rod.  On the left end the $\phi$ cycle degenerates, and on the right end the $\psi$ cycle degenerates.  Thus every 1-cycle is contractible and so the 3-manifold must be $S^3$.  Indeed, this is the Schwarzschild-Tangherlini black hole.  To arrive at Myers-Perry one need only add a spatial component to $v_1$.

\subsection{Building a lens space with rods}

What if the symplectic product of $v_0$ and $v_2$ is not equal to $\pm 1$?  Consider for example the black hole solution
\beq
v_0=(0,1,0)\hsp v_1=(1,0,0)\hsp v_2=(0,-1,p). \label{lensvec}
\eeq   
Now the symplectic product is equal to $p$.  Again the finite rod is an interval on which neither of the two circles degenerate, so the horizon is a 2-torus fibered over an interval.  However at the left end $\phi$ degenerates and at the right end $\phi+p\psi$ degenerates.  These two circles do not generate all of the first homology of the torus $\H^1(T^2)=\Z^2$, but only those elements which wrap the $\psi$ circle a number of times which is divisible by $p$.  Therefore only an index $p$ subgroup of the first homology group of the torus is contractible on the horizon, and so the first homology of the horizon is
\beq
\H_1(\rm{horizon})=\frac{\H_1(T^2)}{\rm{contractible\ cycles}}=\frac{\Z\times\Z}{\Z\times p\Z}=\frac{\Z}{\Z}\times\frac{\Z}{p\Z}=\Z_p.
\eeq
Thus the horizon is not a 3-sphere, which would have been simply connected.  

We will now argue that the horizon has the topology of the lens space
\beq
L(p,1)=S^3/\Z_p. \label{lente}
\eeq
Consider the unit 3-sphere in $\C^2$ with coordinates $z_1$ and $z_2$.  The norm of $z_1$ is always between 0 and 1
\beq
0\leq |z_1|\leq 1
\eeq
and so we will let it parametrize the position on the middle rod.  Let $\phi$ and $\psi$ be the phases of $z_1$ and $z_2$ respectively.  At the left end of the rod, $z_1=0$ and so the $\phi$ circle degenerates, corresponding to a rod $v_0=(0,1,0)$ and at the right end $z_2=0$ and so the $\psi$ circle degenerates, corresponding to $v_2=(0,0,1)$.  So far we have described the 3-sphere in Eq.~(\ref{sfera}).  

So what about the lens space (\ref{lente})?  This is the quotient of the 3-sphere by the free $\Z_p$ action which simultaneously rotates both $z_1$ and $z_2$ by $e^{2\pi i/p}$.  This identification acts on our torus
\beq
(\phi,\psi)\longrightarrow(\phi+2\pi/p,\psi+2\pi/p). \label{ident}
\eeq
While the quotiented 2-torus is still a 2-torus, the $\phi$ and $\psi$ circles no longer generate its' first homology.  There are new cycles in the quotiented space.  For example, you may proceed a distance $2\pi/p$ in the $\phi$ direction, then do one transformation (\ref{ident}) backwards, reducing both $\phi$ and $\psi$ by $2\pi/p$ and then proceed forwards another $2\pi/p$ in the $\psi$ direction.  Then one arrives precisely where one started on the quotiented torus, and so one has described a loop that exists on the quotiented torus but not on the original torus.  Call this new loop $\psi\p$.  While $\phi$ and $\psi$ do not generate the full first homology of the quotiented torus, $\phi$ and $\psi\p$ do generate the full first homology.  Thus we may re-express our rod vectors in the $(\phi,\psi\p)$ basis, which is related to the old $(\phi,\psi)$ basis by
\beq
\phi=\phi\hsp \psi=p\psi\p-\phi.
\eeq
Applying this transformation, one may rewrite the rod vectors (\ref{sfera}) in the basis of the new quotiented torus.  One arrives precisely at Eq.~(\ref{lensvec}).  Therefore the rod structure (\ref{lensvec}) describes a horizon with topology $L(p,1)$.  

If we take the length of the middle rod to zero, then the two semi-infinite rods become adjacent and the lens space shrinks to a point.  A point which is linked by a lens space $L(p,1)$ is called a $\Z_p$ orbifold singularity or a $A_{p-1}$ singularity.  Thus if two adjacent rods have a symplectic product equal to $\pm p$, then their intersection is a $A_{p-1}$ orbifold singularity.

We may generalize Eq.~(\ref{lensvec}) by setting
\beq
v_2=(0,-q,p)
\eeq
and the same argument leads to the identification of the horizon topology as $L(p,q)$.  Thus all of the horizon topologies demonstrated to be consistent in Ref.~\cite{class} are in fact realized by rod diagrams.

However we are not yet finished, because the asymptotics are determined by the semi-infinite rods.  The fact that they have a symplectic product that is not equal to $\pm 1$, implies that the asymptotic region is also $L(p,1)$.  We are interested in producing discrete hair, and so we want an asymptotic region which is a 3-sphere.  Therefore the semi-infinite rods must generate the entire torus, as in the Tangherlini solution (\ref{sfera}).  We can combine the lens space horizon of (\ref{lensvec}) and the asymptotics of (\ref{sfera}) by introducing a fourth rod.  Consider the four rods with vectors
\beq
v_0=(0,1,0)\hsp v_1=(1,0,0)\hsp v_2=(0,-1,p)\hsp v_3=(0,0,1). \label{lensfera}
\eeq   
The $v_3$ rod does not touch the $v_1$ rod, and so it does not affect the horizon topology.  Likewise the $v_1$ and $v_2$ rods do not extend to infinity, and do not affect the asymptotics.  Therefore (\ref{lensfera}) is the rod structure of a black hole with a lens space topology horizon but which asymptotes to unquotiented Minkowski space.  Some examples are well-known.  For example, $p=0$ yields $L_{0,1}=S^2\times S^1$ and so is the nonrotating black ring.  This solution becomes nonsingular only if a spatial component is added to $v_1$ to make it spin sufficiently quickly.  Also in our case angular momentum may be required to eliminate naked singularities, although in our case the singularity will not extend all of the way to the horizon and so may be excised without affecting the horizon topology.  Next, $p=1$ is apparently $L_{1,1}=S^3$ like the Tangherlini black hole.  As no such black holes are expected, except for Tangherlini itself, it is possible that either the conical singularity at $v_2$ cannot be resolved in this case, or else that when it is resolved one finds the Tangherlini solution.  When $p=2$ the horizon topology is $L_{2,1}=\rp^3$, the group manifold of SO(3), this would already be a new solution.  The infinite $p$ limit is the Tangherlini black hole. 

Given the rod structure, the results of Ref.~\cite{Harmark} may be used to obtain the corresponding solution of Einstein's equation from a 3-dimensional matrix valued Poisson type equation sourced by the rods.  When the rod vectors are orthogonal the matrices are diagonal and so the system reduces to 3 decoupled Poisson equations which are easily solved.  If we could write down the solution in the case (\ref{lensfera}) we would similarly be done.  The fact that the symplectic product of $v_2$ and $v_3$ is equal to $-1$ implies that there is no orbifold singularity at their intersection.  If we could fix the lengths of the rods or the rotation of the hole, corresponding to the spatial part of $v_1$ such that there is no conical singularity on $v_2$, then we would have a satisfactory solution which may even be free of naked singularities.  Unfortunately, I don't know how to solve the matrix-valued Poisson's equation even in this simple case, however an interested reader may be able to solve it at least numerically and so vastly improve our results.

For a given rod structure, it is possible that all solutions to Einstein's equations are singular.  For example, the rod structure (\ref{lensfera}), as well as the rod structure (\ref{rodnostri}) to which we will turn momentarily, describe asymptotically Minkowski static solutions of general relativity with horizons of lens space topology.  In Ref.~\cite{nogo} the authors claim that all such solutions must be Tangherlini black holes with topology $S^3$.  In fact they show that the embedding of the horizon must be totally umbilical, which is a local condition and so does not discriminate between spheres and their quotients, and then argue that the only such 3-manifold that can be embedded in $\R^4$ is the $S^3$.  However our black hole is not necessarily embedded in $\R^4$, we impose only that it be embedded in a Ricci flat space which asymptotes to a round $S^3$.  

A necessary condition for such an embedding is that there exist a 4-manifold whose boundaries are the lens space at the horizon, and the $S^3$ at infinity.  When such a 4-manifold exists, one says that the lens space is cobordant to $S^3$.  In fact, the lens space and $S^3$ are cobordant.  To see this, consider multicentered Taub-NUT.  At large radius this asymptotes to a lens space.  Now cut out a small contractible $S^3$ anywhere.  The remaining space is a cobordism between the cut $S^3$ and the asymptotic lens space.  Therefore there is no topological obstruction, there exist 4-manifolds that interpolate between a lens space and $S^3$.  In the Taub-NUT example, the manifold is even Ricci flat, although it interpolates between a small $S^3$ and a large lens space.  In our case we want instead a Ricci flat manifold that interpolates between a small lens space and a large $S^3$.

\subsection{Inverse scattering}
In the present note we will find 5-rod solutions to the 5-dimensional Einstein's equations with vectors
\beq
v_0=(0,1,0)\hsp v_1=(0,-p,1)\hsp v_2=(1,0,0)\hsp v_3=(0,1,p)\hsp v_4=(0,0,1). \label{rodnostri}
\eeq   
where $p$ is an integer strictly greater than one.  The symplectic product of $v_1$ and $v_3$ is equal to $p^2+1$ and so the middle rod corresponds to a horizon with topology $L(p^2+1,1)$.  In particular we do not find all of the possible topologies described above.  The symplectic products of the adjacent vectors $v_0$ and $v_1$ and also $v_3$ and $v_4$ are both equal to one and so there are no orbifold singularities.  We will find in Sec.~\ref{parasec} that for $p>1$ it is possible to choose the positions of the junctions ${a_1,a_2,a_3,a_4}$ of the rods so that there is no conical singularity at $v_1$ and $v_3$.  The absence of a conical singularity at $v_0$ and $v_4$ is easily imposed by fixing the periodicities of the $\phi$ and $\psi$ coordinates to be $2\pi$.  We begin with four free parameters, the points $a_i$.  The elimination of each of the two conical singularities places a single constraint, an overall shift in the position is irrelevant, and so a single parameter remains, which is an overall scale and corresponds to the mass of the solution.  Thus static black lenses are completely characterized by $p$ and their mass, leading to a single countably infinite kind of hair.

We will arrive at these solutions using the inverse scattering technique \cite{Belinsky,Belinsky2,Pomeransky}, via a strategy that roughly mirrors Ref.~\cite{saturn}.  The idea is that we consider configurations with one timelike isometry and two spatial isometries.  Therefore the metric only depends on 2 coordinates.  Einstein's equations are thus reduced to a well-studied 2-dimensional integrable system.  In this system there is a solution generating technique, known as the inverse scattering transformation, which takes a given solution together with some parameters and generates another solution.   Thus we will begin with a solution in which the rod vectors are orthogonal, so that we can easily find that metric, and we will transform it to (\ref{rodnostri}).

The inverse scattering transform consists of two steps.  In the first, one removes solitons from the 2-dimensional integrable system, in the second, one reinserts the same solitons.  Like the rods, each soliton comes with a projective vector, called a BZ vector, in the 3-dimensional space of isometries.  If one reinserts a soliton with the same vector with which is was removed, then one simply returns to the old solution.  The inverse scattering technique is useful because if a soliton is reinserted with a different vector, then one arrives at a new solution.

An arbitrary inverse scattering transformation leads to a singular spacetime.  In order to avoid singularities, solitons may only be removed from and added to the points $a_i$ where the rods $v_{i-1}$ and $v_i$ meet.  Furthermore, both the BZ vector of the soliton removed and the BZ vector of the soliton added must be in the 2-dimensional subspace spanned by the vectors $v_{i-1}$ and $v_i$.  As the length of the BZ vector is irrelevant, the choice of vector contains a single degree of freedom.  In practice it is difficult to perform the transform if the BZ vector of the removed soliton does not lie along either $v_{i-1}$ or $v_i$, and so it does not contribute a degree of freedom.  Thus, for each soliton removed and added, one obtains one additional degree of freedom.  We will be interested in a 2-soliton transform, and so we will introduce two degrees of freedom.  We will find that the asymptotic Minkowski condition imposes one constraint, and the choice of $p$ imposes a second, and so in the end there will be no additional continuous degrees of freedom.

A precise relation between the BZ vectors chosen in an inverse scattering transformation and the resulting rotation of the rod vectors is unknown.  Currently one guesses a set of BZ vectors, tries the transformations and then computes the resulting rod structure using the inverse scattering technique, only at the end learning what the final rod vectors are.  However, intuitively the removal and addition of a soliton at the point $a_i$ rotates all of the rods, even those far away, in the $(v_{i-1},v_i)$ plane, each rod by a different amount which depends on how far it lies from the soliton.  We are only able to solve Poisson's equations for orthogonal rod vectors, and yet we want our inverse soliton transformation to result in the rod structure (\ref{rodnostri}) which is nondiagonal in the $(\phi,\psi)$ plane.  Therefore, we want to create a rotation on the $(\phi,\psi)$ plane, and so will subtract and add solitons at the intersections of rods with vectors $\phi$ and $\psi$.

\section{The ans\"atz} \label{asec}

\subsection{The seed solution}

When the rod vectors are orthogonal, we may immediately write down a solution to Einstein's equations.  Thus we will begin with an orthogonal version of (\ref{rodnostri}), called the seed, with rod vectors
\beq
v_0=(0,1,0)\hsp v_1=(0,0,1)\hsp v_2=(1,0,0)\hsp v_3=(0,1,0)\hsp v_4=(0,0,1) \label{diag}
\eeq   
and rods ending at the points $\{a_1,a_2,a_3,a_4\}$.  The middle rod corresponds to a black hole, but with horizon topology $L_{1,1}=S^3$.  There are no orbifold singularities.  Our goal is to rotate the vectors $v_1$ and $v_3$ into the form (\ref{rodnostri}) via a two-soliton inverse scattering transformation.  As we wish to rotate them on the $(\phi,\psi)$ plane, we will need to remove and introduce solitons at the interface between $\phi$ and $\psi$ rods.  There are two such points, one at $z=a_1$ and one at $z=a_4$, and we will see that solitons must be removed and added at both, so that their effects at infinity may cancel. 

Recall that $z$ is the coordinate along the rods, and $\rho$ is the perpendicular coordinate.  The metric may be written \cite{Harmark}
\beq
ds^2=g_{tt} dt^2+g_{\phi\phi}d\phi^2+2g_{\phi\psi}d\phi d\psi+g_{\psi\psi}d\psi^2+e^{2\nu}(d\rho^2+dz^2). \label{met0}
\eeq
To write down the solution for a given rod structure, it will be convenient to introduce the following notation
\beq
\mu_i=\sqrt{\rho^2+(z-a_i)^2}-(z-a_i). \label{mu}
\eeq
For a diagonal rod structure, each component of the metric $g_{vv}$ in the isometry directions is the product of the $\mu_i$'s for each $i$ such that $a_i$ is the left end of a rod with vector $v$ divided by all of the $\mu_i$'s such that $a_i$ is the right end.  For the rod which is semi-infinite on the left, one uses $\rho^2$ for the left end, while nothing is added for the right end of the rod which is semi-infinite on the right.  To make the metric Minkowski, one simply multiplies $g_{tt}$ by $-1$.  Therefore the metric in the rod directions corresponding to the rod structure (\ref{diag}) is
\beq
(g^{0}_{tt},g^{0}_{\phi\phi},g^{0}_{\psi\psi})=\left(-\frac{\mu_2}{\mu_3},\frac{\rho^2\mu_3}{\mu_1\mu_4},\frac{\mu_1\mu_4}{\mu_2}\right).
\eeq
The zero is used to label the solution before both the soliton removal and the soliton addition.  The metric component $e^{2\nu}$ is, in the diagonal case, the product of a constant of integration $k^2$ by the metric in the direction of the right semi-infinite rod.  This is then multiplied by a product of $R_{ij}$'s, one for each pair ${\mu_i,\mu_j}$ such that $\mu_i$ and $\mu_j$ are on opposite sides of the same component of $g_{vv}$ in the isometry part of the metric, divided by all of the $R_{ij}$'s such that $\mu_i$ and $\mu_j$ are on the same side.  As each $\mu_i$ is on the same side as itself, one obtains the product of all of the $R_{ii}$'s, but each $R_{ii}$ counts for two self pairs, and so the $R_{ii}$'s are not squared.  Therefore, in our case
\beq
e^{2\nu_0}=k^2\frac{\mu_1\mu_4}{\mu_2}\frac{R_{12}R_{13}R_{23}R_{24}R_{34}}{R_{14}^2R_{11}R_{22}R_{33}R_{44}}
\eeq
and we have completely determined the seed metric (\ref{met0}).

\subsection{The inverse scattering transformation}

Now that we have obtained the seed solution, we will modify it by removing two solitons.  We will remove one soliton from $a_1$ and one from $a_4$, both with BZ vector $(0,1,0)$.  Removing a soliton from $a_i$ with vector $v$ corresponds to multiplying the metric component $g_{vv}$ by $-\mu_i^2/\rho^2$.  Therefore we arrive at the new metric
\beq
(g^{0}\p_{tt},g^{0}\p_{\phi\phi},g^{0}\p_{\psi\psi})=\left(-\frac{\mu_2}{\mu_3},\frac{\mu_1\mu_3\mu_4}{\rho^2},\frac{\mu_1\mu_4}{\mu_2}\right).
\eeq
Notice that only the ${\phi\phi}$ component has changed, because both solitons were oriented in the $\phi$ direction.  We will not compute the new $e^{2\nu}$, it turns out that it is much easier to compute this directly at the end, once we have reintroduced the new solitons with new BZ vectors.

We want to put back our solitons at $a_1$ and $a_4$ with new BZ vectors.  The problem is that for a general choice of vectors, the transformation of metric components that depend on $\mu_1$ and $\mu_4$ will be singular.  As $g_{\phi\phi}$ and $g_{\psi\psi}$ are proportional to $\mu_1$ and $\mu_4$, we are unable to reinsert the solitons.  Fortunately, the addition of a soliton commutes with the multiplication of the metric by a function.  Therefore one may divide out the $\mu_1$ and $\mu_4$ dependence, add the solitons, and then put it back in thus avoiding the singular intermediate steps.  In our case, we will divide the metric by $\mu_1\mu_4$, yielding
\beq
(\tilde{g}^{0}_{tt},\tilde{g}^{0}_{\phi\phi},\tilde{g}^{0}_{\psi\psi})=\left(-\frac{\mu_2}{\mu_1\mu_3\mu_4},\frac{\mu_3}{\rho^2},\frac{1}{\mu_2}\right). 
\eeq
The time component now depends on $\mu_1$ and $\mu_4$, however it is unaffected by the addition of the soliton and so this will be irrelevant.  It could be important if one attempts to introduce yet more solitons to make the lens space rotate.

To reinsert solitons with arbitrary BZ vectors, we will need to introduce a new matrix $\Psi(\lambda)$, called the generating matrix, which is a function of the parameter $\lambda$ such that $\Psi(0)$ is equal to the metric of the 3-dimensional space of isometries, and which solves a certain differential equation.  When the isometry components of the metric $\tilde{g}$ are functions only of $\mu_i$ and $\rho^2/\mu_i$, the matrix $\Psi(\lambda)$ is easy to obtain.  One simply substitutes each $\mu_i$ by $\mu_i-\lambda$ and each $\rho^2/\mu_i$ by $\rho^2/\mu_i+\lambda$.  Thus our generating matrix is
\beq
\Psi(\lambda)=\rm{diag}\left(-\frac{\mu_2-\lambda}{(\mu_1-\lambda)(\mu_3-\lambda)(\mu_4-\lambda)},\frac{1}{\frac{\rho^2}{\mu_3}+\lambda},\frac{1}{\mu_2-\lambda}\right). \label{gen}
\eeq
This matrix will provide a $\rho$ and $z$ dependent scaling function for our BZ vectors.

Finally we are ready to reinsert our two solitons.  The first is reinserted at $z=a_1$ with bare BZ vector $m_0^{(1)}$ and the second at $z=a_4$ with bare BZ vector $m_0^{(4)}$ where
\beq
m_0^{(1)}=(0,b_1,c_1)\hsp m_0^{(4)}=(0,b_4,c_4).
\eeq
Only the directions of the BZ vectors affect the final metric, therefore it will be a useful consistency check to see that all calculated quantities are independent of the overall scale of each vector.  The normalized BZ vectors $m^{(k)}$ are the bare vectors $m_0^{(k)}$ scaled by the inverse of the corresponding generating matrix evaluated at $\mu_k$, in other words
\beq
m^{(k)}=m_0^{(k)}\Psi^{-1}(\mu_k). \label{norm}
\eeq
Now we see why it was critical that $g_{\phi\phi}$ and $g_{\psi\psi}$ be independent of $\mu_1$ and $\mu_4$.  Otherwise $\Psi^{-1}$ would have been contained infinite matrix elements which would have scaled some of the components of the BZ vectors to infinity.  $\Psi^{-1}_{tt}(\mu_k)$ is now infinite, but since the time components of the BZ vectors vanish, this matrix element does not contribute to the normalized BZ vectors $m^{(k)}$.  

Substituting the generating matrix Eq.~(\ref{gen}) into the definition (\ref{norm}) one obtains the scaled BZ vectors $m^{(k)}$.  As their time components are equal to zero, we will write only their $\phi$ and $\psi$ components
\beq
m^{(k)}=(b_k\Psi^{-1}_{\phi\phi}(\mu_k),c_k\Psi^{-1}_{\psi\psi}(\mu_k))=\left(\frac{b_kR_{3k}}{\mu_3},c_kD_{2k}\right)=\left(\frac{\hat{b}_k}{\mu_3},\hat{c}_k\right)\hsp k={1,4}
\eeq
where we have introduced the compact notation
\beq
D_{ij}=\mu_i-\mu_j\hsp R_{ij}=\rho^2+\mu_i\mu_j\hsp \hat{b}_k=b_kR_{3k}\hsp \hat{c}_k=c_kD_{2k}. \label{dr}
\eeq

To define the transformed metric, one introduces a two by two symmetric matrix $\Gamma_{kl}$, where $k$ and $l$ label the solitons, and so run over the set $\{1,4\}$.  The matrix is equal to
\beq
\Gamma_{kl}=\frac{m^{(k)}\tilde{g}_0m^{(l)}}{R_{kl}}=\frac{1}{R_{kl}}\left(\frac{\hat{b}_k\hat{b}_l}{\rho^2\mu_3}+\frac{\hat{c}_k\hat{c}_l}{\mu_2}\right). \label{gamma}
\eeq
The matrix $\Gamma$ is then used to construct the final form of the metric.  Putting back the factor of $\mu_1\mu_4$ which was divided out above, the final form of the metric in the isometry directions is
\beq
g_{ab}=\mu_1\mu_4\tilde{g}^0_{ab}-\mu_1\mu_4\sum_{k,l\in\{1,4\}}\frac{\tilde{g}^0_{ac}m_c^{(k)}\Gamma^{-1}_{kl}m_d^{(l)}\tilde{g}_{db}^0}{\mu_k\mu_l}. \label{isomet}
\eeq
Notice that the metric depends on the inverse of $\Gamma$.  This is divergent when $\Gamma$ is degenerate somewhere on the $\rho-z$ plane, which in fact occurs in our case and leads to our naked singularities.  In fact, generically on a two-dimensional plane each eigenvalue will have some zeroes and so we can expect singularities.  In the case of a two-soliton transformation, the determinant consists of only two terms
\beq
\rm{Det}(\Gamma)=\Gamma_{11}\Gamma_{44}-\Gamma_{14}^2 \label{detdef}
\eeq
which vanishes whenever the two terms cancel.  The second term is minus a square and so is negative, thus there is a danger when the first term is positive.

In the particular case in which the transformation mixes a timelike and a spacelike direction it is sometimes possible to arrange the signs such that $\Gamma_{11}\Gamma_{44}$ is negative and so $\Gamma$ is negative definite and nondegenerate.  This occurs in soliton transformations to impart angular momentum along a single axis, as in the black Saturn solution of Ref.~\cite{saturn}.  However more general solutions, with angular momenta along multiple axes as in the bi-ring, more than two solitons, or rotations among spacelike directions like the present case are prone to such degeneracies.  

In our case, in which there are two solitons which mix only spacelike directions, $\Gamma_{11}\Gamma_{44}$ is positive and $-\Gamma_{14}^2$ is negative, thus one may expect at least a coordinate singularity.  As the sign of $g$ changes as the determinant passes zero, past the singularity there are also closed timelike curves.  However fortunately in our case the singularity is contractible and is far from both the black hole and the asymptotic region, in particular there are no closed timelike curves near the black hole, and we will see that its' thermodynamic quantities are well-behaved.  In particular, the ADM mass of our solution is just the Komar mass of the horizon, the singularity does not contribute.  If one deforms the solution so that the determinant is decreased a bit locally near the singularity but away from the hole and the horizon, then the singularity vanishes but one continues to have a lens space horizon and asymptotically flat space.  One may hope that as in the case of the black ring either adding angular momentum to the black lens, or quantum effects or the effects of accreted material near the singularity, may lead to such a deformation.

To finish specifying the metric (\ref{met0}), we need now only provide $e^{2\nu}$.  This is given by a particularly simple formula.  If $\Gamma_0$ is equal to $\Gamma$ when the old BZ vectors are reinserted, {\it{i.e.}} at $b_1=b_4=1$ and $c_1=c_4=0$ then
\beq
e^{2\nu}=e^{2\nu_0}\frac{\rm{det}(\Gamma)}{\rm{det}(\Gamma_0)}. \label{rhomet}
\eeq
Now in principle the metric (\ref{met0}) is completely determined.  $\mu_i$ are functions of $\rho$, $z$ and the parameters $a_i$ via Eq.~(\ref{mu}).  $D_{ij}$, $R_{ij}$, $b^{(k)}$ and $c^{(k)}$  are functions of the $\mu_i$ via Eq.~(\ref{dr}).  The matrix $\Gamma$ is given in Eq.~(\ref{gamma}) and the metric components in Eqs.~(\ref{isomet}) and (\ref{rhomet}) are given in terms of these.  

\subsection{Rewriting the solution}

To analyze the solution, it will be convenient to substitute in Eq.~(\ref{gamma}), explicitly taking the inverse of $\Gamma$.  The inverse of a symmetric two by two matrix is given simply by exchanging the two diagonal elements, negating the off-diagonal elements and dividing by the determinant.  Therefore the determinant of $\Gamma$ will appear in the denominator of the angular components of the metric, as it appears in the numerator of the $\rho$ and $z$ components by Eq.~(\ref{rhomet}).  Therefore it will be convenient to factor out Det($\Gamma$) from the metric components
\beq
g_{\phi\phi}=\frac{A_{\phi\phi}}{\rm{Det}(\Gamma)}\hsp
g_{\pi\psi}=\frac{A_{\phi\psi}}{\rm{Det}(\Gamma)}\hsp
g_{\psi\psi}=\frac{A_{\psi\psi}}{\rm{Det}(\Gamma)}. \label{gang}
\eeq

The determinant of $\Gamma$ is determined by inserting (\ref{gamma}) in (\ref{detdef}).  After a bit of rearrangement, it is
\beq
\rm{Det}(\Gamma)=\frac{1}{R_{11}R_{44}}\left[\frac{(\bo\cfo-\bfo\co)^2}{\rho^2\mu_2\mu_3}-\frac{\rho^2D_{14}^2}{R_{14}^2}\left(\frac{\bo\bfo}{\rho^2\mu_3}+\frac{\co\cfo}{\mu_2}\right)^2\right]. \label{detgamma}
\eeq
Inverting the matrix $\Gamma$ using the explicit form (\ref{detgamma}) of the determinant one finds the numerators of the angular metric components of (\ref{gang})
\bea
A_{\phi\phi}&=&-\frac{1}{\mu_1\mu_4R_{11}R_{44}}\left[\frac{1}{\rho^2\mu_2}(\bo\cfo\mu_4-\bfo\co\mu_1)^2+\frac{D_{14}^2}{\mu_3R_{14}^2}\left(\bo\bfo-\co\cfo\frac{\mu_1\mu_3\mu_4}{\mu_2}\right)^2\right]\\
A_{\psi\psi}&=&-\frac{1}{\mu_1\mu_2\mu_3\mu_4R_{11}R_{44}}\left[\frac{1}{\mu_2}(\bo\cfo\mu_1-\bfo\co\mu_4)^2+\frac{\mu_1^2\mu_4^2D_{14}^2}{\rho^2\mu_3R_{14}^2}\left(\bo\bfo-\co\cfo\frac{\rho^4\mu_3}{\mu_1\mu_2\mu_4}\right)^2\right]\nonumber\\
A_{\phi\psi}&=&\frac{D_{14}}{\mu_1\mu_2\mu_4R_{11}R_{14}R_{44}}\left[\mu_4R_{11}\bo\co\left(\frac{\bfo^2}{\rho^2\mu_3}+\frac{\cfo^2}{\mu_2}\right)-\mu_1R_{44}\bfo\cfo\left(\frac{\bo^2}{\rho^2\mu_3}+\frac{\co^2}{\mu_2}\right)\right].\nonumber \label{ansatz}
\eea
Using Eq.~(\ref{rhomet}) one may also express $e^{2\nu}$ in terms of the hatted BZ vectors.  Now we need to determine the parameters.

\section{Choosing the parameters} \label{parasec}
Eq.~(\ref{ansatz}) is our ans\"atz.  It contains the parameters $a_1,\ a_2,\ a_3,\ a_4,\ b_1,\ b_4,\ c_1$\ and $c_4$.  An overall shift in the $z$ axis is irrelevant, so we are only interested in the shift independent combinations
\beq
K_{ij}=a_i-a_j.
\eeq
Furthermore the overall scale of each BZ vector is irrelevant, so the metric will only depend on the combinations $c_1/b_1$ and $c_4/b_4$.  This leaves five parameters, of which three are independent $K_{ij}$'s and two are ratios $c/b$.  There is also the constant of integration $k$ which appeared in $e^{2\nu_0}$.  We are also free to determine the periodicities of the coordinates $\phi$ and $\psi$.  With these parameters we want to impose that the spacetime is asymptotically Minkowski, that there are no conical singularities on the spacelike rods $a_1<z<a_2$ and $a_3<z<a_4$ and that the rod vectors are indeed (\ref{rodnostri}).  We will see in the rest of this section that these conditions use up all of the degrees of freedom of the parameters except for an overall scale, which determines the mass of the black lens.  In particular no parameters will be left to eliminate the naked singularity.

\subsection{Asymptopia} \label{asysec}
The novelty of our black lens solution is that the asymptotic spacetime is globally Minkowski space.  To study the asymptotic region, we define the radial coordinate $r$ by
\beq
\rho=r\sin(2\theta)\hsp z=r\cos(2\theta).
\eeq
Now we may expand everything in powers of $1/r$.  To leading order
\beq
\rho\sim 2r\sin^2(\theta)\hsp D_{ij}\sim 2K_{ij}\sin^2(\theta)\hsp R_{ij}\sim 4r^2\sin^2(\theta) 
\eeq
which determine the hatted BZ vectors
\beq
\hat{b}_k\sim 4b_k\sin^2(\theta)\hsp \hat{c}_k\sim 2c_kK_{2k}\sin^2(\theta).
\eeq
Substituting this limiting behavior into the ans\"atz (\ref{detgamma}) and (\ref{ansatz}), nearly everything becomes independent of the $c$'s
\beq
\rm{Det}(\Gamma)\sim -\frac{b_1^2b_4^2K_{41}^2}{4r^4\sin^2(\theta)\cos^2(\theta)}\hsp
A_{\phi\phi}\sim -\frac{b_1^2b_4^2K_{41}^2}{2r^3\sin^2(\theta)}\hsp
A_{\psi\psi}\sim -\frac{b_1^2b_4^2K_{41}^2}{2r^3\cos^2(\theta)}.
\eeq
In particular the ratios $g_{\phi\phi}$ and $g_{\psi\psi}$ tend to $2r\cos^2(\theta)$ and $2r\sin^2(\theta)$ respectively.  If one defines the radial coordinate to be $\sqrt{2r}$, this produces the usual form of $g_{\phi\phi}$ and $g_{\psi\psi}$ in Cartesian coordinates for Minkowski space, if they have periods $2\pi$.

In Cartesian coordinates $g_{\psi\phi}$ is zero.  However, in the large $r$ limit $g_{\phi\psi}$ grows linearly in $r$.  The term linear in $r$ is proportional to a particular combination of the BZ vectors
\beq
g_{\phi\psi}\propto (b_4c_1K_{21}+b_1c_4K_{42})r + \mathcal{O}(r^{-1}). \label{asyrot}
\eeq
In other words, not only do the $\phi$ and $\psi$ coordinates successfully mix where they need to near the horizon, but also they continue to rotate among each other as one proceeds out to infinity, instead of separating as in Minkowski space.  Fortunately the rotation is determined by an expression in (\ref{asyrot}) which is independent of $\rho$ and $z$, and so may be eliminated everywhere with a single constraint on the parameters.  This constraint, which we will impose from now on, is
\beq
b_4c_1K_{21}+b_1c_4K_{42}=0. \label{stella}
\eeq
Once it is imposed, $g_{\phi\psi}$ falls as $1/r$ and so eventually tends to zero as $r$ tends to infinity, as it must if the asymptotic space is to be Minkowski.  Intuitively the condition (\ref{stella}) can be interpreted as follows.  If one thinks of $c_i/b_i$ as a force exerted on the $(\phi,\psi)$ plane at the point $a_i$ then the left hand side of (\ref{stella}) is the torque about the axis $a_2$.  The balancing condition then implies that the torque about $a_2$ must vanish.  If instead of subtracting solitons with bare BZ vectors in the $\phi$ direction we had chosen solitons with BZ vectors in the $\psi$ direction then it would have been the torque at $a_3$ which would need to vanish.  If the constraint (\ref{stella}) is not imposed, one may still arrive at the usual Minkowski coordinates asymptotically by a nonorthogonal rotation of $\phi$ and $\psi$.  One would then need to impose that the rotated coordinates are $2\pi$-periodic instead of $\psi$ and $\phi$ and one may repeat our analysis, hoping that the additional degree of freedom allows one to escape the naked singularity.

So far we have seen that the angular part of the metric agrees with Minkowski space with one restriction on our parameters.  Next we will test the $\rho$ and $z$ parts of the metric, which are given by a single function $e^{2\nu}$, which is calculated using Eq.~(\ref{rhomet}) from $e^{2\nu_0}$, Det($\Gamma$) and Det($\Gamma_0$).  The leading behavior of these in $r$ is
\beq
\rm{Det}(\Gamma_0)\sim -\frac{K_{41}^2}{4r^4\sin^2(\theta)\cos^2(\theta)}\hsp
e^{2\nu_0}\sim \frac{k^2}{2r}.
\eeq
Therefore
\beq
g_{\rho\rho}=g_{zz}=e^{2\nu}\sim \frac{k^2b_1^2b_4^2}{2r}.
\eeq
Demanding that, with radial coordinate $\sqrt{2r}$ one arrives in Cartesian coordinates fixes the constant of integration $k$ to be
\beq
k=\frac{1}{b_1b_4} \label{inconst}
\eeq
where we have made the substitution 
\beq
\frac{dr^2}{2r}=(d\sqrt{2r})^2
\eeq
in the line element squared (\ref{ansatz}).

The last nontrivial component of the metric that needs to be checked is $g_{tt}$.  This component was unchanged by the inverse scattering transformation and so it still asymptotes to $-1$.  However its simple form allows one to easily expand it to the next order
\beq
g_{tt}\sim -1+\frac{K_{32}}{r} \label{tdep}
\eeq
and so we see in Sec.~\ref{propsec} that the ADM mass of our solution will be proportional to $K_{32}$.  We will see later that it is just equal to $3\pi K_{32}$, independently of the choices of the other parameters.  But before calculating the ADM mass, we will nonetheless fix the other parameters.

\subsection{Conical singularities}

In fixing the asymptotic structure we have put a single constraint on the three independent $K_{ij}$'s and two ratios $c/b$, we have fixed the constant of integration $k$, and we have fixed the periods of $\phi$ and $\psi$.  We therefore have four remaining parameters.  As the periods of $\phi$ and $\psi$ have been fixed, the elimination of conical singularities on the spatial singularities will require a further fixing of two parameters, one at each rod, leaving one parameter to determine $p$ and one to determine the ADM mass. We will now find the conditions under which the conical singularities are eliminated.

First consider the spacelike rod $a_1<z<a_2$.  Near this rod it will be convenient to introduce the set of dependent positive coordinates
\beq
z_1=z-a_1\hsp z_i=a_i-z
\eeq
where $i$ runs over the values $2$, $3$ and $4$.  The distance from the rod is parametrized by $\rho$ and so we will be interested in an expansion of the metric in powers of $\rho$.  To leading order in $\rho$ we find
\bea
&&\mu_1\sim \frac{\rho^2}{2z_1}\hsp
\mu_i\sim 2z_i+\frac{\rho^2}{2z_i}\hsp
D_{ij}\sim 2K_{ij}\hsp
D_{i1}\sim 2z_i \nonumber\\
&&R_{ij}\sim 4z_iz_j\hsp
R_{i1}\sim\rho^2\frac{K_{i1}}{z_1}\hsp
R_{11}\sim\rho^2.
\eea
Using the balancing condition (\ref{stella}) to eliminate $c_1$, one finds that the hatted BZ vectors near the rod tend to
\beq
\bo\sim b_1\rho^2\frac{K_{31}}{z_1}\hsp
\bfo\sim 4b_4z_3z_4\hsp
\co\sim -2b_1\frac{c_4}{b_4}\frac{K_{42}}{K_{21}}z_2\hsp
\cfo\sim -2c_4K_{42}.
\eeq
One may then substitute these limits into Eqs.~(\ref{detgamma}) and (\ref{ansatz}) to determine the leading behavior of the metric.

Unlike the large $r$ limit, there are multiple terms at leading order in most of the metric components.  This is fortunate.  In the large $r$ limit there is only one term because the soliton transformation parameters $c$ are irrelevant asymptotically, where the spacetime assumes the original Minkowski form.  However while we want the the asymptotic form to rest Minkowski, the soliton transformation needs to transform the horizon topology from a 3-sphere to a lens space.  Therefore it is critical that the $c$ dependence not drop out near the black hole.

Keeping only the leading terms we find
\bea
\rm{Det}{(\Gamma)}&\sim&\frac{4b_1^2c_4^2}{\rho^4}\left[\frac{K_{42}^2}{K_{21}^2}z_2z_3\frac{b_4^2}{c_4^2}-\left(\frac{K_{31}}{K_{41}}\frac{b_4}{c_4}z_4+\frac{K_{42}^2}{K_{21}K_{41}}\frac{c_4}{b_4}z_1\right)^2\right]\nonumber\\
A_{\phi\phi}&\sim&-\frac{8b_1^2c_4^2}{\rho^4}z_1z_3z_4\left(\frac{K_{31}}{K_{41}}\frac{b_4}{c_4}-\frac{K_{42}^2}{K_{21}K_{41}}\frac{c_4}{b_4}\right)^2\nonumber\\
A_{\phi\psi}&\sim&-\frac{8b_1^2c_4^2}{\rho^4}z_1z_3z_4\frac{K_{42}}{K_{21}}\left(\frac{K_{31}}{K_{41}}\frac{b_4}{c_4}-\frac{K_{42}^2}{K_{21}K_{41}}\frac{c_4}{b_4}\right)\nonumber\\
A_{\psi\psi}&\sim&-\frac{8b_1^2c_4^2}{\rho^4}z_1z_3z_4\frac{K_{42}^2}{K_{21}^2}.
\eea
Both the $A$'s and the determinant scale as $\rho^{-4}$ and so there is no divergence at the rod.  However as $\rho$ goes to zero, the $A$ matrix becomes degenerate.  The zero eigenvector is the rod vector
\beq
v_1=(0,1,\frac{K_{21}K_{31}}{K_{41}K_{42}}\frac{b_4}{c_4}-\frac{K_{42}}{K_{41}}\frac{c_4}{b_4}). \label{v1}
\eeq
To agree with the desired rod structure (\ref{rodnostri}) we will need the combination of parameters on the right to be equal to $-1/p$.  We will impose this condition in Subsec.~\ref{topsec}.

For now we will try to understand the conical singularity at this rod.  The $v_1$ circle degenerates at $\rho=0$ and its circumference grows linearly with the radius as one leaves the rod.  To avoid a conical singularity, one needs to impose that at small $\rho$ the circumference is equal to $2\pi$ times the radius.  In other words, one needs to calculate the next to leading order contribution to the metric along this circle, which will be nonzero and will be of order $\rho^2$, and to fix its ratio with respect to the metric $e^{2\nu}$ in the $\rho$ direction.  We know that the period of the $\phi$ and $\psi$ circles is $2\pi$, but the $v_1$ circle is neither of these, it is a combination.  It must be a rational combination, or else it will never close and there will inevitably be a conical singularity.  Thus we will find a discrete condition.  We will fix the period later when we impose that $v_1$ corresponds to (\ref{rodnostri}), for now we solve for the period as a function of the parameters.

The period of the isometry in the $v_1$ direction, in order to avoid a conical singularity, must be equal to the small $\rho$ limit of
\beq
\Delta v_1=2\pi\sqrt{\frac{\rho^2 e^{2\nu}}{g_{v_1v_1}}}=2\pi\sqrt{\frac{\rho^2e^{2\nu_0}\rm{Det}(\Gamma)^2}{A_{v_1v_1}\rm{Det}(\Gamma_0)}}. \label{debito}
\eeq
While $e^{2\nu_0}$ is independent of $c/b$, the determinant of $\Gamma$ has different powers of $c/b$ with distinct $z$ and $\rho$ dependences ranging from degree 0 to degree 4.  Furthermore the determinant is squared in (\ref{debito}), and so there are really coefficients from degree 0 to 8.  The individual components of $g$ only have components over a range of four powers of $c/b$, so naively it seems impossible for the period (\ref{debito}) to be independent of $\rho$ and $z$. 

The situation is saved by the fact that $g_{v_1v_1}$ is not a component of $g$ in the original basis.  Instead it is the length squared of $v_1$, which has a term of order $b/c$ and one of order $c/b$. Therefore when squared it gives contributions of order $b^2/c^2$ to $c^2/b^2$ to $g$, which already had terms of order $1$ to $c^4/b^4$.  Therefore in the end the numerator of (\ref{debito}) has terms of order $0$ to $8$ in $c/b$, and the denominator has terms of order $-2$ to $6$.  Therefore it is possible that the ratio is fixed, and furthermore the period will be proportional to the square root of $c^2/b^2$, or in other words it will be proportional to $c/b$.  In particular we will see that at $c=0$ there will be a singularity unless the rod length goes to zero, reducing our solution to Tangherlini.  

As the validity of the inverse scattering transform guarantees that the period of the $v_1$ circle will be $z$-independent, it suffices to insert just one order in $c$ in Eq.~(\ref{debito}).  We will consider the $c^6$ term in the dominator, which we have argued corresponds to the $c^8$ term in the numerator.  This term is convenient to consider because it has only a single contribution in the numerator and in the denominator, both of which occur at the leading order of their respective terms in the small $\rho$ expansion.   However we have checked that the same condition is reproduced by the $c^{-2}$ term in the denominator, which corresponds to the $c^0$ term in the numerator.

In the numerator of (\ref{debito}), the only contribution at order $c^8$ comes from the square of the $c^4$ term in Det$(\Gamma)$, which to leading order is
\beq
\rm{Det}(\Gamma)\sim -\frac{4}{\rho^4}\frac{b_1^2c_4^4}{b_4^2}\frac{K_{42}^2}{K_{21}^2K_{41}^2}z_1^2.
\eeq
In the denominator of (\ref{debito}) the only contribution at order $c^6$ comes from the $c^4$ term in $A_{\psi\psi}$ multiplied by the $c^2$ term in the expansion
\beq
A_{v_1v_1}=v_1Av_1=A_{\phi\phi}+2v_1^\psi A_{\phi\psi}+(v_1^{\psi})^{2}A_{\psi\psi}
\eeq
where $v_1^\psi$ is given in Eq.~(\ref{v1}).  The leading contribution to the $c^4$ term of $A_{\psi\psi}$ is
\beq
A_{\psi\psi}\sim -\frac{2}{\rho^2}\frac{b_1^2c_4^4}{b_4^2}\frac{z_1^3}{z_2z_4}\frac{K_{42}^4}{K_{21}^2K_{41}^2}.
\eeq
Finally the $c$-independent factors at leading order are
\beq
e^{2\nu_0}\sim \frac{1}{2b_1^2b_4^2}\frac{K_{21}K_{31}}{K_{41}^2}\frac{z_4}{z_1z_2}\hsp
\rm{Det}(\Gamma_0)\sim -\frac{4}{\rho^4}\frac{K_{31}^2}{K_{41}^2}z_4^2.
\eeq
Substituting all of these expressions into (\ref{debito}) one arrives at the deficit angle
\beq
\Delta v_1=2\pi\frac{c_4}{b_4}\frac{K_{42}}{\sqrt{K_{21}K_{31}}}. \label{per1}
\eeq
As expected, it is linear in $c$.  Also it is invariant under an overall scaling of the $a_i$'s, which only changes of the mass of the black lens.  It is also invariant under a rescaling of the BZ vectors, which leaves $c_4/b_4$ constant, and under a translation of the $a_i$'s.  Most importantly it is independent of $z$, it is constant along the rod and so it will lead to a single condition that can be imposed on the parameters, once we solve for the left hand side in Subsec.~\ref{topsec}.

The same analysis may be applied to the rod which extends from $a_3$ to $a_4$.  Now the small $\rho$ behavior of the $\mu$'s is slightly different.  If we define
\beq
z_i=z-a_i \hsp z_4=a_4-z
\eeq
where $i$ runs over the values $1$, $2$ and $3$, then to leading order
\bea
&&\mu_i\sim \frac{\rho^2}{2z_i}\hsp
\mu_4\sim 2z_4+\frac{\rho^2}{2z_4}\hsp
D_{ij}\sim \frac{\rho^2K_{ij}}{2z_iz_j}\hsp
D_{4i}\sim 2z_4 \nonumber\\
&&R_{ij}\sim \rho^2\hsp
R_{4i}\sim\rho^2\frac{K_{4i}}{z_i}\hsp
R_{44}\sim 4z_4^2.
\eea
Substituting these approximations into (\ref{ansatz}) we again find that the angular components of the metric are degenerate at small $\rho$.  This time the zero eigenvector is equal to
\beq
v_3=(0,-\frac{K_{43}}{K_{41}}\frac{b_4}{c_4}+\frac{K_{42}}{K_{41}}\frac{c_4}{b_4},1). \label{v3}
\eeq
Later we will impose that this is proportional to the value in Eq.~(\ref{rodnostri}).

The conical singularity condition for $\Delta v_3$ is also given by Eq.~(\ref{debito}).  The powers of $c$ that appear in the numerator and denominator are identical to those of the other rod.  This time we consider the $c^0$ term in the numerator and $c^{-2}$ term in the denominator, which comes from the product of the $c^0$ term in $g_{\phi\phi}$ with the $c^{-2}$ term in $(v_3^{\phi})^2$.  The leading $\rho$ behavior of the factors in (\ref{debito}) that contribute at these orders in $c$ are
\bea
&&e^{2\nu_0}\sim \frac{1}{2b_1^2b_4^2}\frac{K_{42}K_{43}}{K_{41}^2}\frac{z_1}{z_3z_4}\hsp
\rm{Det}(\Gamma)\sim -\frac{4}{\rho^4}b_1^2b_4^2\frac{K_{43}^2}{K_{41}^2}z_1^2\\
&&\rm{Det}(\Gamma_0)\sim -\frac{4}{\rho^4}\frac{K_{43}^2}{K_{41}^2}z_1^2\hsp
A_{v_3v_3}\sim-\frac{2}{\rho^2}\frac{K_{43}^4}{K_{41}^2}\frac{z_1^3}{z_3z_4}\frac{b_1^2b_4^4}{c_4^2}.\nonumber
\eea
Inserting all of these factors in (\ref{debito}) we find the periodicity condition for the $v_3$ isometry
\beq
\Delta v_3=2\pi\frac{c_4}{b_4}\frac{\sqrt{K_{42}}}{\sqrt{K_{43}}}. \label{per3}
\eeq
Equations (\ref{per1}) and (\ref{per3}) now give the periodicities of the cycles in the $v_1$ and $v_3$ directions.  When they are satisfied, the configuration is free of conical singularities.

\subsection{Horizon topology}  \label{topsec}
The periods of $v_1$ and $v_3$ are already determined.  There is only one 2-torus, generated by $\phi$ and $\psi$, whose periods are each $2\pi$.  The $v_1$ and $v_3$ cycles are on this torus, and so their lengths are determined from $v_1$ and $v_3$.  In this subsection we will impose that $v_1$ and $v_3$ are, up to an irrelevant scale, those of Eq.~(\ref{rodnostri}), so that our horizon is the lens space $L(p^2+1,1)$.  This leads to four new constraints.  

In order to avoid a conical singularity at $a_1$ and $a_4$, the $v_1$ and $v_3$ cycles need to wrap the $\psi$ and $\phi$ cycles respectively just once, although they can wrap the $\phi$ and $\psi$ cycles an arbitrary integral number of times.  From (\ref{rodnostri}) we see that $v_1$ wraps the $\psi$ cycle once and the $\phi$ cycle $-p$ times, while $v_3$ wraps the $\phi$ cycle once and the $\psi$ cycle $p$ times.  As $v_1^\phi=1$, the fact that $v_1$ wraps the $\phi$ cycle $-p$ times implies that its period is $q$ times that of the $\phi$ cycle, in other words
\beq
\Delta v_1=2\pi p.
\eeq
Of course, only the absolute value of the period is fixed by the conical singularity condition.  Similarly $v_3^\psi=1$ and so the fact that $v_3$ wraps the $\psi$ cycle $p$ times implies that its period is $p$ times that of the $\psi$ cycle
\beq
\Delta v_3=2\pi p.
\eeq
Inserting these two values in Eqs.~(\ref{per1}) and (\ref{per3}) one obtains the first two constraints.  The second two constraints are just that $v_1$ and $v_3$ are proportional to the cycles that they wrap.  As we have normalized $v_1^\phi=v_3^\phi=1$, the other components are then
\beq
v_1^\psi=-\frac{1}{p}\hsp v_3^\phi=\frac{1}{p}.
\eeq
Inserting these relations into Eqs.~(\ref{v1}) and (\ref{v3}) one obtains the other two constraints.

Therefore conical singularity free lens spaces $L(p^2+1,1)$ are described by solutions of four equations.  If we introduce the shorthand notation and normalization
\beq
c=\frac{c_4}{b_4}\hsp
K_{41}=1\hsp y=K_{42}\hsp x=K_{43}\hsp 1-y=K_{21}\hsp 1-x=K_{31}
\eeq
then these equations are
\beq
|c\sqrt{\frac{y}{x}}|=p\hsp
|c\frac{y}{\sqrt{(1-x)(1-y)}}|=p\hsp
\frac{1}{p}=-\frac{x}{c}+cy\hsp
-\frac{1}{p}=\frac{(1-x)(1-y)}{yc}-yc.
\eeq
Here we have taken $p$ to be positive, corresponding to a choice of orientation for the angular coordinates.  As $x$ and $y$ are positive, the absolute values may be moved onto $c$ alone.  For a given value of $p$ these are 4 equations for 3 variables, and so it is not clear that a solution exists.  We will be saved by a $\Z_2$ symmetry in the system of equations which inverts $z$, exchanging $\phi$ with $\psi$ and $x$ with $y$.

The right hand sides of the first two constraints are equal, so their left hand sides are equal.  Dividing through by $c$ this implies
\beq
\sqrt{\frac{y}{x}}=\frac{y}{\sqrt{(1-x)(1-y)}}.
\eeq
If we multiply both sides by $x$ and then square them
\beq
xy=(1-x)(1-y)=(1-x)-y+xy
\eeq
and so
\beq
y=1-x. \label{yeq}
\eeq
Therefore the rods are symmetric with respect to $z$ reflections and we can eliminate $y$.
We can now use the first constraint to eliminate $c$
\beq
|c|=p\sqrt{\frac{x}{1-x}}. \label{cequ}
\eeq

Now we can use Eqs.~(\ref{yeq}) and (\ref{cequ}) to express the third constraint entirely in terms of $x$ and $p$.  Taking $c$ to be positive
\beq
\frac{1}{p}=-\frac{\sqrt{x(1-x)}}{p}+p\sqrt{x(1-x)}=\sqrt{x(1-x)}\left(p-\frac{1}{p}\right).
\eeq
This is easily solved for $x(1-x)$
\beq
x(1-x)=\frac{1}{(p^2-1)^2}
\eeq
which is a quadratic equation for $x$ with two roots.  As we want the horizon to have a positive size, we will take the root for which $x<1/2$
\beq
x=\frac{1-\sqrt{1-4/(p^2-1)^2}}{2}\hsp y=\frac{1+\sqrt{1-4/(p^2-1)^2}}{2}.
\eeq
In particular, we have no solution at $p=1$ as $x$ is infinite, but for $p>1$ we find that $0<x<1/2$ and so all rods have finite, positive length.  In particular, the square rooted quantities are positive, and so there is a solution for every $p>1$.  Re-expressing this result in our old notation 
\beq
\frac{K_{43}}{K_{41}}=\frac{1-\sqrt{1-4/(p^2-1)^2}}{2}\hsp \frac{K_{42}}{K_{41}}=\frac{1+\sqrt{1-4/(p^2-1)^2}}{2}
\eeq
and Eq.~(\ref{cequ}) yields the BZ vector
\beq
\frac{c_4}{b_4}=p\sqrt{\frac{1-\sqrt{1-4/(p^2-1)^2}}{1+\sqrt{1-4/(p^2-1)^2}}}.
\eeq
Recall that the other BZ vectors were determined in terms of these by Eq.~(\ref{stella})
\beq
\frac{c_1}{b_1}=-p\sqrt{\frac{1+\sqrt{1-4/(p^2-1)^2}}{1-\sqrt{1-4/(p^2-1)^2}}}.
\eeq
 With these relations, we have completely determined the metric as a function of $p$ and an overall scale, for example $K_{41}$.  There is no constraint on the value of the scale, all of the relations are homogeneous. 

\section{Physical properties} \label{propsec}
Now that we have found our black lens, we turn to studying its properties.  In particular we want to know the Komar mass of the horizon, its entropy, its temperature and its ADM mass.  The first 3 quantities are found by analyzing the metric near the horizon.  In other words we take a small $\rho$ limit similar to that done near the spacelike rods.

Again, we begin by defining the degenerate coordinates
\beq
z_i=z-a_i \hsp z_m=a_m-z
\eeq
where $i$ and later $j$ run over the values $1$ and $2$ and $m$ and later $n$ run over $3$ and $4$.  To leading order
\bea
&&\mu_i\sim \frac{\rho^2}{2z_i}\hsp
\mu_m\sim 2z_m+\frac{\rho^2}{2z_m}\hsp
D_{ij}\sim \frac{\rho^2K_{ij}}{2z_iz_j}\hsp
D_{mi}\sim 2z_m\hsp
D_{mn}\sim 2K_{mn} \nonumber\\
&&R_{ij}\sim \rho^2\hsp
R_{mi}\sim\rho^2\frac{K_{mi}}{z_i}\hsp
R_{mn}\sim 4z_mz_n.
\eea
which lead to the following hatted BZ vectors
\beq
\bo\sim b_1\rho^2\frac{K_{31}}{z_1}\hsp
\bfo\sim 4b_4z_3z_4\hsp
\co\sim -\frac{b_1c_4}{2b_4}K_{42}\frac{\rho^2}{z_1z_2}\hsp
\cfo\sim -2c_4z_4.
\eeq

Substituting these limits into (\ref{ansatz}), without imposing that the conical singularities vanish but imposing the balancing relation (\ref{stella}) between $c_1$ and $c_4$, we find the leading order behavior of the various components of the metric
\bea
\rm{Det}(\Gamma)&\sim&-\frac{4}{\rho^4}\frac{z_4^2}{K_{41}^2}(b_1b_4K_{31}+\frac{b_1}{b_4}c_4^2K_{42})^2 \label{hor}\\
A_{\phi\phi}&\sim& -\frac{8b_1^2c_4^2}{\rho^4}\left[\frac{z_3z_4}{K_{41}^2z_1}(K_{31}\frac{b_4}{c_4}z_1-K_{42}\frac{c_4}{b_4}z_4)^2+\frac{K_{31}^2z_2z_4}{z_1}\right]\nonumber\\
A_{\psi\psi}&\sim& -\frac{8b_1^2c_4^2}{\rho^4}\left[\frac{z_2z_4}{K_{41}^2z_1}(K_{31}\frac{b_4}{c_4}z_4-K_{42}\frac{c_4}{b_4}z_1)^2+K_{42}^2\frac{z_3z_4}{z_1}\right]\nonumber\\
A_{\phi\psi}&\sim& \frac{8b_1^2c_4^2}{\rho^4}\frac{z_4}{K_{41}z_1}\left[(K_{31}^2z_2z_4-K_{31}K_{42}z_1z_3)\frac{b_4}{c_4}+(K_{42}^2z_3z_4-K_{42}K_{31}z_1z_2)\frac{c_4}{b_4}\right]\nonumber\\
e^{2\nu_0}&\sim&\frac{1}{2b_1^2b_4^2}\frac{K_{31}K_{32}K_{42}}{K_{41}^2}\frac{1}{z_2z_3}.\nonumber 
\eea
First we will use these components to calculate the volume of the horizon.

The horizon is at $\rho=0$, $a_2\leq z\leq a_3$ and so it extends along the $\phi$, $\psi$ and $z$ directions.  Therefore a volume element is the square root of the determinant of the metric in the $\phi$, $\psi$ and $z$ directions.  As in the black Saturn case, this element is independent of $z$ and so the volume is simply the product of the ranges of the coordinates by the square root of the determinant.  The ranges of $\phi$ and $\psi$ are each $2\pi$, while that of $z$ is $K_{32}$, therefore the volume is
\beq
\rm{Vol}=4\pi^2K_{32}\sqrt{\rm{Det}(g)}=4\pi^2K_{32}\sqrt{e^{2\nu}}\sqrt{g_{\phi\phi}g_{\psi\psi}-g_{\phi\psi}^2}=4\pi^2K_{32}e^{2\nu_0}\sqrt{\frac{A_{\phi\phi}A_{\psi\psi}-A_{\phi\psi}^2}{\rm{Det}(\Gamma)\rm{Det}(\Gamma_0)}}.
\eeq
Substituting in the leading order behaviors (\ref{hor}) we find
\beq
\rm{Vol}=\pi^2\sqrt{32}\frac{K_{32}}{K_{41}}\sqrt{\frac{K_{32}K_{42}}{K_{31}}}(K_{31}+\frac{c_4^2}{b_4^2}K_{42}).
\eeq

If we define the length scale
\beq
K_{41}=2L^2
\eeq
then, after removing the conical singularities, the volume is
\beq
\rm{Vol}=8\pi^2L^3\left(1-\frac{4}{(p^2-1)^2}\right)^{3/4}\left(p^2+1-(p^2-1)\sqrt{1-\frac{4}{(p^2-1)^2}}\right).
\eeq
This expression will simplify momentarily when we fix the scale by setting the mass to one.

To fix the temperature $T$ we note that the temporal part of the metric is unchanged, and so
\beq
g_{tt}=-\frac{\mu_2}{\mu_3}\sim -\frac{\rho^2}{4z_2z_3}. \label{gtt}
\eeq
Defining the inverse temperature to be the period of the imaginary time if there is no conical singularity in the complexified spacetime
\beq
\frac{1}{T}=2\pi\sqrt{-\frac{\rho^2 e^{2\nu}}{g_{tt}}}=2\pi\sqrt{-\frac{\rho^2 e^{2\nu_0}\rm{Det}(\Gamma)}{g_{tt}\rm{Det}(\Gamma_0)}}
\eeq
we find that the temperature is
\beq
T=\frac{K_{41}}{\pi\sqrt{8}}\sqrt\frac{K_{31}}{K_{32}K_{42}}\frac{1}{(K_{31}+\frac{c_4^2}{b_4^2}K_{42})}
\eeq
where again we have not imposed the absence of conical singularities.

The Komar mass of the horizon is
\beq
M=\frac{3}{32\pi}\int*d\xi
\eeq
where $\xi$ is the one-form $dt$ dual to the Killing direction $t$.  As the metric is $t$-independent, Eq.~(\ref{gtt}) implies that $d\xi$ is proportional to $dt\wedge d\rho$.  Therefore Hodge duality multiplies it by the square root of the determinant of the components of the metric in the other directions, that is, in the $z$ direction and in the angular directions.  The integral of the determinant in the $z$ and angular directions just gives the volume of the horizon.  Thus the integral yields the volume of the horizon times the magnitude of $dt$ in units of the $t$ and $\rho$ part of the volume form, which is just the $\rho$ derivative of $g_{tt}$ normalized by $1/\sqrt{g_{\rho\rho}g_{tt}}$.  As $g_{tt}$ is quadratic is $\rho$
\beq
\partial_\rho g_{tt}=2g_{tt}/\rho
\eeq
which when divided by $\sqrt{g_{\rho\rho}g_{tt}}$ just gives the temperature times a constant.  More precisely
\beq
M=\frac{3}{32\pi}K_{32}4\pi^2\sqrt{g_{zz}(g_{\phi\phi}g_{\psi\psi}-g_{\phi\psi}^2)}\frac{\partial_\rho g_{tt}}{\sqrt{g_{\rho\rho}{g_{tt}}}}=\frac{3}{8}\rm{Vol}\cdot T=3\pi K_{32}=6\pi L^2\sqrt{1-\frac{4}{(p^2-1)^2}}.
\eeq
and so black lenses satisfy the Smarr relation.

We can set the mass to one by setting
\beq
L=\frac{1}{\sqrt{6\pi}}\left(1-\frac{4}{(p^2-1)}\right)^{-1/4}
\eeq
which loses no interesting physics as all formulas are homogeneous in $L$.  We can now calculate the entropy of the mass one black lens with horizon $L(p^2+1,1)$.  It is
\beq
S=\frac{\rm{Vol}}{4}=\frac{1}{3}\sqrt{\frac{\phi}{6}}(p^2+1-\sqrt{(p^2-1)^2-4}).
\eeq
Normalizing the dimensionless area $a_H$ as in Ref.~\cite{saturn}
\beq
a_H=\frac{3}{16}\sqrt{\frac{3}{\pi}}A=\frac{1}{4\sqrt{2}}(p^2+1-\sqrt{(p^2-1)^2-4}).
\eeq
In particular at $p=2$, describing the lens space $L(5,1)$ the dimensionless area is about $.48$, which is much less than $2\sqrt{2}$, the entropy of the corresponding Tangherlini black hole.  At higher $p$ the entropy continues to decrease, asymptoting to $1/\sqrt{8}$.  Therefore it is likely that black lenses are classically unstable and decay into Tangherlini black holes, although they may be metastable as in the black Saturns of Ref.~\cite{metastable}.  Also black lenses of higher $p$ may decay to black lenses of lower $p$, or when they collide they may result in products at lower $p$.

To calculate the ADM mass one uses the same procedure as for the calculation of the Komar mass at the horizon, but instead of the small $\rho$ limit one uses the large $r$ limit described in Subsec.~\ref{asysec}.  There the asymptotics of the relevant quantities have already been provided.  Now one needs to use the second term in the expansion (\ref{tdep}), as the leading term is killed by the exterior derivative.  The integral is now done over a three-sphere at large $r$.  It gives the same answer as the integral over the event horizon above.  And so the ADM mass is equal to the Komar mass.  In particular, there is no net mass between the horizon and infinity.

\section{Conclusions}

In this note we have used the inverse scattering technique to find a family of new static solutions of Einstein's equations.  They are black holes whose event horizons have topology $L(p^2+1,1)$ and which are asymptotically Minkowski.  To our knowledge this is the first time that black holes with lens space event horizons have been embedded in a spacetime which is not asymptotically a quotient of Minkowski space or a multi-centered Taub-NUT space.  We have numerically calculated the Ricci tensor and we have seen that it vanishes everywhere that we have checked within the precision of Mathematica.  These solutions contain no free parameters other than their total mass and the topology of the lens space.  However we have constructed solutions for a countably infinite number of topologies, and so we have provided a new countably infinite variety of hair for 5-dimensional black holes.

Unfortunately our solutions appear to have two spherical naked singularities, located between the horizon and infinity but whose interiors do not contain the black hole.  Inside of the naked singularities there are closed timelike curves.  As one approaches the singularity the square of the Riemann tensor appears to suffer an $r^{-6}$ divergence.  Therefore these solutions appear to be pathological.  On the positive side, the singularity does not approach the black hole, unlike the conical singularity in the case of the nonrotating black ring, although it does surround the points $z=a_1$ and $z=a_4$ at $\rho=0$.  There are two reasons to suspect that it may be possible to eliminate this singularity.  First, it does not contribute to the total ADM mass.  Therefore its elimination does not require any changes in the asymptotic structure, it can be done locally, without even affecting the horizon topology.  Secondly if one does deform it away, say by decreasing the determinant of $\Gamma$ near the singularity, the remaining space is nonsingular.  In other words, the singularity is not necessary to support the fact that the topology of the horizon is a lens space.  

One natural candidate for a deformation that would eliminate the singularity is an angular momentum for the black lens.  After all, the black lens does, like the black ring, contain a noncontractible cycle and it may be that, as is the case for the conical singularity of the black ring, the tension of this cycle is somehow responsible for the singularity.  The black lens is not a black ring, the rest of the black lens geometry will exert a force on the cycle and so could stabilize it, but perhaps this force is too strong?  In the Tangherlini $L(1,1)$ case the force is just right to avoid a singularity.  Another possible deformation would occur in an embedding in a theory of quantum gravity.  Perhaps, as in the G\"odel case studied by Gimon and Ho\v{r}ava, some quantum degrees of freedom condense near the singularity and cut it out or replace it.  In this case, as the singularity was not necessary for the topology of the horizon or for the asymptotic flatness, we would have again a black lens in an asymptotically Minkowski space.  Therefore it may be interesting to embed this solution in a quantum completion such as string theory, or to add the non-Lorentz invariant counterterms resulting from RG flow in a background with a big bang singularity which may render gravity renormalizable, at least in 4d as in Refs.~\cite{Damiano,Damiano2}.  Finally it may be that adding some other fields to the theory may eliminate the singularity, as they may accrete near it and screen or transform it.  The simplest way to proceed would be to numerically search for generalizations, for example solving directly the matrix valued Poisson equation for rod structures with angular momenta.

It may be that while only spherical horizon black holes exist in asymptotically Minkowski space, $L(p,q)$-horizons exist when the asymptotic geometry is $L(r,s)$ with $r>p$.  Therefore discrete hair may exist in asymptotically taub-NUT space for example.

While this solution itself is pathological, we may draw two conclusions.  First, inverse scattering transformations are often degenerate and this can lead to naked singularities, particularly when one does something more complicated than adding angular momentum in a single plane via a 2-soliton transformation.  Solutions such as the bi-ring should be checked for such singularities.  Second, solutions that appear to require a particular kind of asymptotic structure may be consistent with a different asymptotic structure at yet larger scales.  The rod formalism seems to particularly adapted to such embeddings, as changing the physics at larger scales corresponds to simply adding additional rods at the ends.

\section* {Acknowledgement}

\noindent
I have benefited from the insight of R. Emparan, D. Klemm, C. Krishnan, A. Magni and D. Persson, as well as from the funding of SISSA and its ideal beach for relaxing calculations.


\end{document}